\documentclass[12pt]{amsproc}
\usepackage{amsmath,amssymb}
\usepackage{pstricks}
\usepackage{pst-all}

\usepackage[T1]{fontenc}

\title[An urn model associated with the Jacobi polynomials]{An urn model associated with the Jacobi polynomials}
\author{F. Alberto Gr\"unbaum}
\thanks{The author was supported in part by NSF Grant DMS-0603901.}
\date{}
\address{Department of Mathematics \\ University of California \\ Berkeley,
CA\ \ 94720}

\begin{document}

\begin{abstract}

We consider an urn model leading to a random walk that can be solved explicitly in terms
of the well known Jacobi polynomials.
\end{abstract}

\maketitle

\bigskip
\bigskip
\bigskip

\section{Urn models and orthogonal polynomials}

There are two simple and classical models in statistical mechanics which
have recently been associated to very important classes of orthogonal polynomials. The oldest one of this models is due to 
D. Bernoulli (1770) and S. Laplace (1810), while the more recent model is due
to Paul and Tatiana Ehrenfest (1907). While both of these models are featured in very classical texts in probability theory, such as \cite{F}, the connection
with orthogonal polynomials is of much more recent vintage. In fact, the polynomials in question due to Krawtchouk and Hahn had not been recognized as basic
objects with rich properties till around 1950. For a few pertinent and useful references see \cite{A,AAR,Ch,DS,E,G1,ILMV,K,KMcG1}.  

From the comments above one could get the impression that the relations between orthogonal polynomials -specially some well known classes of them- is only a matter of historical interest. Nothing could be further from the truth: there are several areas of probability and mathematical physics where recent important progress hinges on the connections with orthogonal polynomials.

The entire area of Random Matrix theory starts with the work of E. Wigner and F. Dyson and reaches a new stage in the hands of M. Mehta who brought in the power of orthogonal polynomials into the picture.  

In the area of Random growth processes the seminal work of K. Johansson depends heavily on orthogonal polynomials , specifically Laguerre and Meixner ones, in \cite{J}.

The connection between birth-and-death processes and orthogonal polynomials has many parents but the people that made the most of it are S. Karlin and J. McGregor, see \cite{KMcG}. We will have a chance to go back to their work in connection with our model here. One should remark that the ideas of using the spectral
analysis of the corresponding one step transition matrix have been pushed recently in the case of quantum random walks, an area where physics, computer science and mathematics could make important contributions. See \cite{CGMV,Ko}.

The study of the so called ASEP (asymmetric simple exclusion processes) going back to F. Spitzer, see \cite{S},  and very much connected with the work of K. Johansson mentioned earlier. has recently profited from connections with the Askey-Wilson polynomials. All of this has important and deep connections with combinatorics and a host of other areas of mathematics.
There are many other examples that one could mention, but we just add here the study of nonintersecting or non-colliding random process which goes back to F. Dyson.

There are lots of interrelations among these areas. For one example: the Hahn polynomials that were mentioned in connection with the Bernoulli-Laplace model  were studied by Karlin and McGregor in connection with a model in genetics due to Moran, see \cite{KMcG1}. They have also been found to be useful in discussing random processes with non-intersecting paths, see \cite{G}.

All of these areas are places where orthogonal polynomials have been put to very good use. For a very good review of several of these items see \cite{K}. Orthogonal polynomials of several variables, as well as matrix valued orthogonal polynomials have recently been connected to certain random walks. For two samples see \cite{G2,G3}.

\bigskip

\section{The Jacobi polynomials}

The classical Jacobi polynomials are usually considered either in the
interval $[-1,1]$ or, as will do, in the interval $[0,1]$

These polynomials are orthogonal with respect to the weight function

$$
W(x)=x^{\alpha}(1-x)^{\beta}.$$

\noindent
Here we assume that $\alpha, \beta>-1$,
in fact it will be assumed throughout that $\alpha , \beta$ are non-negative
integers.

These polynomials are eigenfunctions of the differential operator

$$x(1-x)\frac{d^2}{dx^2}+(\alpha+1+x(\alpha+\beta+2))\frac{d}{dx}
$$

\noindent
a fact that will not play any role in our discussion but which is crucial
in most of the physical and/or geometrical applications of Jacobi polynomials.
These applications cover a vast spectrum including potential theory, electromagnetism and
quantum mechanics.

Neither the orthogonality, nor the fact that our polynomials are eigenfunctions of this differential operator are enough to determine them uniquely. One can
multiply each polynomial by a constant and preserve these properties. We chose
to normalize our polynomials by the condition $$Q_n(1)=1.$$

For us it will be important that these polynomials satisfy (in fact are defined by) the three term recursion relation

$$xQ_n(x)=A_{n}Q_{n+1}(x)+B_nQ_n(x)+C_{n}Q_{n-1}(x)
$$

with $Q_0=1$ and $Q_{-1}=0$.

\bigskip

The coefficients $A_n,B_n,C_n$ given by

\begin{align*}
    A_n=&\frac{(n+\beta+1)(n+\alpha+\beta+1)}{(2n+\alpha+\beta+1)(2n+\alpha+\beta+2)},\quad n\geq0\\
    B_n=&1+\frac{n(n+\beta)}{2n+\alpha+\beta}-\frac{(n+1)(n+\beta+1)}{2n+\alpha+\beta+2},\quad n\geq0\\
    C_n=&\frac{n(n+\alpha)}{(2n+\alpha+\beta)(2n+\alpha+\beta+1)},\quad
    n\geq1,\\
\end{align*}

The coefficient $B_n$ can be rewritten as
\begin{align*}
   B_n=\frac{2 n (n+\alpha+\beta+1)+(\alpha+1)\beta +\alpha(\alpha+1)}{(2 n +\alpha+\beta)(2 n +\alpha+\beta+2)}
\end{align*}

\noindent
which makes it clear that, along with the other coefficients, it is non-negative.

\bigskip
\bigskip

Since we insist on the condition $Q_n(1)=1$ we can see, for instance by induction and using the recursion relation, that $$A_n+B_n+C_n=1.$$

There are, of course, several explicit expressions for the different variants
of the Jacobi polynomials and they can be used for instance in computing the
integrals that appear in the last section.

\bigskip

The normalization above is natural when one thinks of these polynomials (at least for some values of $\alpha,\beta$) as the spherical functions for some appropriate symmetric space, and insists that these functions take the value $1$ at the North pole of the corresponding sphere. The simplest of all cases is the one with $\alpha=\beta=0$ when one gets the Legendre polynomials and the usual two dimensional sphere sitting in ${\mathbb R}^3$. The reader may want to see \cite{VK}.

The fact that the coefficients are nonnegative and add up to one
cries out for a probabilistic interpretation of these quantities. This is the purpose of this paper. We have not seen
in the literature concrete models of random walks where the Jacobi polynomials play this role.

\section{The model}

Here we consider a discrete time random walk on the non-negative integers whose
one step transition probability matrix coincides with the one that gives the
three-term recursion relation satisfied by the Jacobi polynomials.

\bigskip

At times $t=0,1,2,....$ an urn contains $n$ blue balls and this determines the state
of our random walk on ${\mathbb Z}\geq 0$.

The urn sits in a "bath" consisting of an infinite number of red balls. The transition mechanism
 is made up of a few steps which are described now, leaving some of the details for later.

In the first step a certain number of red balls from the surrounding bath are mixed with the
$n$ blue balls in the urn.

In the second step a ball is selected (with uniform distribution) from among the balls in the urn. This "chosen ball" can be blue or red. In either case an experiment is performed in a parallel world, using an appropriate "auxiliary urn", to determine if this chosen ball will retain its color or have it changed (from red to blue or viceversa).

Once this is settled, and the possible change of color has taken place,  the main urn contains the initial $n$ balls plus a certain number
of balls taken form the bath in the first step, and we are ready for the third and last step. This final step consists of having all red balls in the urn removed and
dropped in the bath.

The state of the system at time $t+1$ is given by the number of blue balls in the urn after these three step are completed. Clearly the new state can take any of the values $n-1,n,n+1$.

A more detailed description of the three steps above is given in the next section.

\section{The details of the model}

If at time $t$ the urn contains $n$ blue balls, with $n=0,1,2,....$ we pick $$n+\alpha+\beta+1$$
red balls from the bath to get a total of $2 n+\alpha+\beta+1$ balls in the urn at the end of step one.

We now perform step two: this gives us a blue ball with probability $$\frac{n}{2 n+\alpha+\beta+1}$$ and a red ball with
probability  $$\frac{n+\alpha+\beta+1}{2 n +\alpha+\beta+1}$$

\bigskip

If the chosen ball is blue then in a "auxiliary urn" with $n$ blue balls we throw in $\alpha$
blue balls and $n+\beta$ read balls, mix all these balls and pick one with uniform distribution. We imagine the auxiliary urn surrounded by a bath of an infinite number of blue and red balls which are used to augment the $n$ blue balls in this auxiliary urn.

The probability of getting a blue ball in the auxiliary urn is $$\frac{n+\alpha}{2 n+\alpha+\beta}$$ and if this is the outcome the "chosen ball" in the main urn has its color changed (from blue to red). If we get a red ball in this auxiliary urn then the chosen ball retains its blue color.

\bigskip

On the other hand, if in step two we had chosen a red ball then in a different "auxiliary urn" with $n$ blue balls we throw in $\alpha+1$ blue balls and $n+\beta+1$ red balls. This auxiliary urn is also surrounded by a bath of an infinite number of blue and red balls.

These balls are mixed and one is
chosen with the uniform distribution. The probability that this ball in the auxiliary urn is red  is given by $$\frac{n+\beta+1}{2 n+\alpha+\beta+2}$$ and if this is the case the chosen ball in the main urn has its color changed (from red to blue). Otherwise the chosen ball retains its red color. 

\bigskip

Notice that the chosen ball in the main urn has a change of color only when we get a match of colors for the balls drawn in the main and an auxiliary urn: blue followed by blue or red followed by red.

\bigskip

In either case once the possible changes of color of the chosen ball in the main urn has been decided upon, we
remove all the red balls form the main urn.
\bigskip

\bigskip

We see that the state of the system goes from $n$ to $n-1$ when the chosen ball is blue and then its color gets changed into red. This event has probability $$\frac{n}{2 n+\alpha+\beta+1}$$
multiplied by $$\frac{n+\alpha}{2 n+\alpha+\beta}$$

Observe that this coincides with the value of $C_n$ in the recursion relation satisfied by our version of the Jacobi polynomials.

\bigskip

The state increases from $n$ to $n+1$ if the chosen ball is red and its color gets changed into blue. This event has probability $$\frac{n+\alpha+\beta+1}{2 n+\alpha+\beta+1}$$ multiplied by $$\frac{n+\beta+1}{2 n+\alpha+\beta+2}$$

Observe that this coincides with the values of $A_n$ given earlier.

\bigskip

As we noticed earlier, 
when the chosen ball is blue and the ball in the corresponding auxiliary ball is red then the chosen ball retains its color. Likewise if the chosen ball is red and the ball in the corresponding auxiliary urn is blue then the chosen one retains its color. In either case the total number
of blue balls in the main urn remains unchanged and the state goes from $n$ to $n$.

Recall the basic property of the coefficients $A_n,B_n,C_n$, namely

$$A_n+B_n+C_n=1 $$

This shows that the probabilty of going from state $n$ to state $n$ is given
by $B_n$.

\bigskip

In summary we have built a random walk whose one step transition probability is the
one given by the three term relation satisfied by our version of the Jacobi polynomials.

\section{Birth-and-death processes and orthogonal polynomials}

\bigskip

A Markov chain with state space given by the non-negative integers and a tridiagonal one step transition probability matrix ${\mathbb P}$ is called a birth-and-death process. Our model given above, clearly fits in this framework.

One of the most important connections between orthogonal polynomials and birth-and-death processes, such as the one considered here is given by the Karlin-McGregor formula \cite{KMcG}.

If the polynomials satisfy

\[
\pi_j \int_{0}^1 Q_i(x)Q_j(x) W(x) dx = \delta_{ij}
\]

\noindent
one gets the following
representation formula for the entries of the powers of the one-step transition
probability matrix

\[
({\mathbb P}^n)_{ij} = \pi_j \int_{0}^1 x^nQ_i(x)Q_j(x)W(x) dx.
\]

This compact expression gives the solution to the dynamics of our random walk and allows for the study
of many of its properties.

\bigskip

In the case of our version of the Jacobi polynomials the squares of the norms of the polynomials $Q_i$ are given by

    $$\frac{\Gamma(i+1)\Gamma(i+\alpha+1)\Gamma(\beta+1)^2}{\Gamma(i+\beta+1)\Gamma(i+\alpha+\beta+1)(2i+\alpha+\beta+1)}$$

\bigskip

In our case when $\alpha , \beta$ are assumed to be nonnegative integers this expression can, of course, be written without any reference to the Gamma function.

\bigskip

\bigskip

We close this paper by recalling how one can compute in the case of
our transition 
matrix $P$ its invariant (stationary) distribution,
i.e. the (unique up to scalars) row vector
$$\mbox{\boldmath$\pi$}=(\pi_0,\pi_1,\pi_2,\dots)$$
such that
$$\mbox{\boldmath$\pi$}P = \mbox{\boldmath$\pi$}. $$

It is a simple matter of using the recursion relation for the polynomials $Q_i$ to show that the components $\pi_i$ are given, up
to a common multiplicative constants by the inverses of the integrals

$$\int_{0}^1 Q^2_i(x) W(x)dx.$$

\noindent
mentioned above.
This justifies the notation $\pi_i$ for these two apparently unrelated quantities, and in our case furnishes
an explicit expression for an invariant distribution.


\clearpage

\begin{thebibliography}{KMcG}




\bibitem[A]{A} Askey, R. {\em Evaluation of Sylvester type determinants using orthogonal polynomials}, Advances in Analysis, Proceed. 4th international ISAAC
Congress, ed. H.G.W. Begehr et al, World Scientific, Singapore, (2005), 1--16.

\bibitem[AAR]{AAR} Andrews, G., Askey, R., and Roy, R., {\em Special
functions},
Encyclopedia of Mathematics and its applications, Cambridge University
Press, 1999.

\bibitem[CGMV]{CGMV} Cantero, M.J. , Gr\"unbaum, F. A. , Moral, L. and Velazquez, L., {\em Matrix valued Szeg\H o polynomials and quantum random walks}, arXiv:0901.2244 (2009)


\bibitem[Ch]{Ch} Chihara, T., {\em An introduction to orthogonal polynomials},
Gordon and Breach Science Publishers, 1978.


\bibitem[DS]{DS} Diaconis, P. and Shahshahani, M., {\em Time to reach stationarity in the Bernoulli-Laplace diffusion model}, Siam J. Math. Anal. 18, 208--218, 1987. 



\bibitem[E]{E} Ehrenfest, P. and Eherenfest, T. , {\em \"Uber zwei 
bekannte Einw\"ande gegen das
Boltzmannsche H-Theorem}, Physikalische Zeitschrift, vol 8, 1907, 311--314.

\bibitem[F]{F}  Feller, W., {\em An introduction to Probability
Theory and its applications}, vol 1, 3rd edition, Wiley 1967.


\bibitem[G]{G} Gorin, V. , {\em Non-intersecting paths and Hahn orthogonal polynomial ensemble}, arXiv:0708.2349 17 Aug 2007



\bibitem[G1]{G1} Gr\"unbaum, F. A., {\em
Random walks and orthogonal polynomials: some challenges}, 
in Probability, Geometry and Integrable systems, Mark Pinsky and Bjorn Birnir editors, 
MSRI publication vol 55, 2007, pp. 241--260 , see also arXiv math PR/0703375 .





\bibitem[G2]{G2} Gr\"unbaum, F. A. , {\em The Karlin-McGregor formula for a variant of a discrete version of Walsh's spider}, J. Phys. A, Sep 2009.



\bibitem[G3]{G3} Gr\"unbaum, F. A. , {\em The Rahman polynomials are bispectral} Sigma 3 , (2007) 065. 



\bibitem[ILMV]{ILMV}  Ismail, M.E.H., Letessier, J., Masson, D., and
Valent, G., {\em Birth and death processes
and orthogonal polynomials}, Orthogonal Polynomials, P. Nevai
(editor)  Kluwer Acad. Publishers, 1990.





\bibitem[J]{J} Johansson, K. , {\em Shape fluctuations and Random Matrices}, Comm. Math. Phys. 209, 437--476, (2000)


\bibitem[K]{K} Kac, M., {\em Random wak and the theory of Brownian 
motion}, American Math. Monthly (1947)
54, 369--391.



\bibitem[KMcG]{KMcG}  Karlin, S., 
and McGregor, J., {\em Random walks},
Illinois J. Math. {\bf 3} 
(1959), 66--81.


\bibitem[KMcG1]{KMcG1}  Karlin, S., and McGregor, J., {\em The Hahn
polynomials, formulas and applications}, Scripta Math. 26, 33-46.


\bibitem[Ko]{Ko} Konno, N. , {\em Orthoogonal polynomials induced by discrete-time quantum walks in one dimension}, arXiv 0903.4047 24 Mar 2009.


\bibitem[K]{K}  K\"oning, W. , {\em Orthogonal polynomial ensembles in probability theory}, Probability Surveys, vol. 2 (2005) 385--447.




\bibitem[S]{S} Spitzer, F. , {\em Interaction of Markov processes},  Advances in Mathematics 5, 246--290 (1970).


\bibitem[VK]{VK} Vilenkin, N., and Klimyk, A., {\em Representation of Lie
Groups and Special functions}, vol.3, Kluwer Academic, Dordrecht,
MA, 1992.










\end{thebibliography}
\end{document}